\documentclass{emulateapj}

\newcommand{\percmsq}{cm$^{-2}$}

\newcommand{\kmpers}{km~s$^{-1}$}

\newcommand{\kms}{km~s$^{-1}$}
\newcommand{\HI}{\ion{H}{1}}

\begin{document}

\title{The Role of Pressure in GMC Formation }
\author{Leo Blitz and Erik Rosolowsky}
\affil{Department of Astronomy, University of California at Berkeley, 
\\601 Campbell Hall MC 3411, CA 94720}
\email{blitz@astro.berkeley.edu}
\begin{abstract}

We examine the hypothesis that hydrostatic pressure alone determines
the ratio of atomic to molecular gas averaged over a particular radius
in disk galaxies.  The hypothesis implies that the transition radius,
the location where the ratio is unity, should always occur at the same
value of {\it stellar} surface density in all galaxies.  We examine
data for 28 galaxies and find that the stellar surface density at the
transition radius is indeed constant to 40\% at a value of 120
M$_\sun$ pc$^{-2}$.  If the hypothesis can be confirmed at all radii
within a large range of galaxy types and metallicities, combining it
with the observed constancy of the star formation rate with H$_2$ surface
density may enable a physically motivated star formation prescription
with wide applicability.

\end{abstract}

\section{Introduction}

Current ideas about the formation of GMCs present a nagging puzzle.
Some authors have suggested that GMCs form by agglomeration of
pre-existing molecular clouds (e.g. Scoville \& Hersh 1979; Kwan \&
Valdes 1983).  Others have argued that GMCs form primarily from atomic
gas through some sort of instability or large-scale shock
(e.g. Woodward 1976; Blitz \& Shu 1980; Engargiola, et al. 2003).  In
principle, both modes may occur: in galaxies that are predominantly
atomic, GMCs might form by one process, and in galaxies that are
primarily molecular, the GMCs might form by another.  If so, two
different GMC formation mechanisms could be at work in the same galaxy
because many galaxies show a transition from being predominantly
molecular at their centers to being predominantly atomic in their
outer parts (e.g. Wong \& Blitz 2002; Helfer et al. 2003). For
example, in the outer parts of spiral galaxies there is so little
molecular gas (e.g. Dame et al. 1987; Dame 1993; Heyer et al. 1998)
that making GMCs from atomic gas seems to be the only available
formation pathway.  On the other hand, the centers of many galaxies
are so overwhelmingly molecular (e.g. Mauersberger et al. 1989; Young
et al. 1995; Wong \& Blitz 2002, Helfer et al. 2003) that it is
implausible for the inner galaxy molecular clouds to form from
anything other than pre-existing molecular gas.  The stars in these
galaxies form only from molecular gas, yet there is no obvious change
in the star forming properties across the molecular/atomic transition
region.  How could it be that molecular clouds that form by two
independent processes would show no obvious difference in their star
formation properties?

One possibility is that the process of cloud formation is independent
of whether the preexisting gas is atomic or molecular.  That is, the
ratio of atomic/molecular gas depends on some other factor, such as
the ambient pressure.  Because the rate of star formation depends only
on the amount of molecular gas \citep{2002ApJ...569..157W}, whatever
determines the molecular gas surface density, determines the variation
of star formation within a galaxy.

In this paper, we consider the hypothesis that the molecular gas
fraction in a galaxy disk is determined by the mean hydrostatic
pressure at a particular radius.  We show that if the hydrostatic
pressure is the only parameter determining the molecular gas fraction,
one predicts that the radius at which the atomic and molecular gas
surface densities are the same, the transition radius, occurs at a
constant value of the {\it stellar} surface density, $\Sigma_*$.
Remarkably, in a sample of 28 galaxies, $\Sigma_*$ is found to be
constant to about 40\% at the transition radius, even though the
observed variation of $\Sigma_*$ in these galaxies is at least 3
orders of magnitude.

\section{Background}

\label{background}

Several authors have previously suggested that gas pressure determines
the molecular fraction at a given radius in a galaxy.  Spergel \&
Blitz (1992), for example, argued that the extraordinarily large
molecular gas fraction at the center of the Milky Way is plausibly the
result of the very high hydrostatic pressure in the Galactic bulge.
Elmegreen (1993) suggested on theoretical grounds that the ratio of
atomic to molecular gas in galactic disks results from both the
ambient hydrostatic pressure as well as the mean radiation field.  The
dependence on the pressure is steeper than that of the radiation
density ($f_{mol} \propto P^{2.2} j^{-1.1}$) and ought to be the
dominant factor.  Observationally, Wong and Blitz (2002) showed that
the radial dependence of the atomic to molecular gas ratio in seven
molecule rich galactic disks can be understood to be the result of the
variation in interstellar hydrostatic pressure (with $f_{mol} \propto
P^{0.8}$).  

Let us assume, therefore, that $N$(H$_2$)/$N$(\HI), the ratio of H$_2$
column density to \HI~column density, is determined {\it only} by the midplane
hydrostatic pressure, $P_{ext}$.  In an infinite disk with isothermal
stellar and gas layers, and where the gas scale height is much less
than the stellar scale height, as is typical in disk galaxies,
to first order: 

\begin{equation}
P_{ext} = (2G)^{0.5}\Sigma_g v_g\{{\rho_*}^{0.5} + (\frac{\pi}{4} 
\rho_g)^{0.5}\}     
\label{fullpressure}
\end{equation}

\noindent where $\Sigma_{g}$ is the total surface density of the gas,
$v_g$ is the velocity dispersion of the gas, $\rho_*$ is the midplane
surface density of stars, and $\rho_g$ is the midplane
surface density of gas.  The first term on the right is due to the hydrostatic
pressure of the gas in the stellar potential; the second term is due 
to the self-gravity of the gas.

In most galaxy disks, $\rho_*$ is much larger than $\rho_g$ when
averaged over azimuth, except in
the far outer parts of a galaxy where the stars become quite rare.  In
the solar vicinity, for example, $\rho_*$ = 0.1 M$_{\sun}$ pc$^{-3}$
(e.g. Binney \& Merrifield 1998),
but $\rho_g \simeq$ 0.02 M$_{\sun}$ pc$^{-3}$ (e.g. Dame 1993).
For a self-gravitating stellar disk, $\Sigma_* = 2\sqrt2 \rho_* h_*$, 
where $h_*$ is the stellar scale height and $h_* = ({v_*}^2/4\pi G
\rho_*)^{0.5}$.  Thus, neglecting $\rho_g$,
Equation 1 becomes: 

\begin{equation}
P_{ext} =  0.84 (G \Sigma_*)^{0.5}\Sigma_g \frac {v_g} {(h_*)^{0.5}} \
\label{approxpressure}
\end{equation}
  
\noindent where $\Sigma_{g}$ is the total surface density of the gas,
$v_g$ is the velocity dispersion of the gas, $\rho_*$ is the midplane
surface density of stars, and $\rho_g$ is the midplane surface density
of gas.  Direct solution of the fluid equations by numerical
integration shows that this approximation is accurate to 10\% for
$\Sigma_* > 20~ M_{\odot}\mbox{ pc}^{-2}$ (where $\rho_* \rightarrow
\rho_g$), which covers the range of stellar surface densities in this
study.

We choose to express the midplane pressure in the form of Equation
\ref{approxpressure} since there is good evidence that both $h_*$ and
$v_g$ are constant with radius in disk galaxies.  Furthermore, because
of the weak dependence of $P_{ext}$ on $h_*$, and the small
variation of $h_*$ measured among galactic disks, we expect variations
of $h_*$ to have little effect on $P_{ext}$.  The 
constancy of the stellar scale height within galaxies was demonstrated
by \citet{vdks1,vdks2} and has been confirmed in other
edge-on galaxies \citep[e.g.][]{fry}.  While there is some evidence
that the stellar scale height flares at large radius in some galaxies
\citep{narayan}, this is only found only at the edges of 
stellar disks in regions where the stellar, gaseous and dark matter
components of the disk make comparable contributions to the potential. 
Adopting a constant stellar scale height is further
supported by the observations of \citet{bottema}, who shows
that the stellar velocity dispersion follows an exponential
distribution with scale length twice that of the stellar surface
density in disks.  This observation suggests that the stellar disk is
distributed vertically in a sech$^2 z$ profile with $\sigma_* = (\sqrt2
\pi G \Sigma_* h_*)^{1/2}$.  

While the stellar component of galactic disks
can be approximated with a constant scale height, the gas component is
better described by a constant velocity dispersion,
$\sigma_g$, which is observed in face-on galaxies
\citep[e.g.][]{svdk,dhh90} and the Milky Way \citep{burton,malhotra}.
The Milky Way observations show that the gas scale height decreases as
the stellar surface mass density increases in a manner consistent the
gas remaining isothermal \citep{malhotra}. This ensemble of
observations suggests that $\sigma_g = 8$ km s$^{-1}$ characterizes
the \ion{H}{1} velocity dispersion in the stellar dominated regions of
galactic disks.

By assumption,
\begin{equation}
N(\mbox{H}_2)/N(\mbox{\ion{H}{1}}) = f(P_{ext}) 
\end{equation}

\noindent Thus, since $(v_g/\sqrt {h_*})$ is approximately constant
within galaxies, then 

\begin{equation}
N(\mbox{H}_2)/N(\mbox{\ion{H}{1}}) = 
f\left[P_{ext}(\Sigma_g,\Sigma_*)\right]
\end{equation}

The mass surface density of atomic gas, $\Sigma_{\mathrm{HI}}$, is
reasonably constant across the inner portions of galactic disk 
\citep[e.g.][]{1986AnAS...66..505W,
1994AJ....107.1003C,2002ApJ...569..157W}.  
While some changes of $\Sigma_{\mathrm{HI}}$
with radius are observed, the variation is small compared to
the changes observed in stellar surface mass density and molecular
surface mass density.  We may therefore adopt a single value for the
surface density of atomic gas in a galaxy across the stellar disk.
$N$(\HI) saturates at a value of about $1 \times 10^{21}$ \percmsq
\citep{2002ApJ...569..157W}.
Thus, $N$(H$_2$)/$N$(\HI) has a value of unity when
2$N$(H$_2$)$+N$(\HI) has a value of $\sim 2 \times 10^{21}$ \percmsq.
Equation \ref{approxpressure}, therefore implies that the radius where
the atomic and molecular surface densities are equal in spiral
galaxies depends only on $\Sigma_*$ if $v_g$/$h_*$ is constant.

\section{Results}
\label{results}

We use BIMA SONG \citep{Helfer03}, an interferometric imaging survey
of the CO in 44 nearby spirals, to determine the distribution of the
molecular gas.  To determine the H$_2$ surface density, we use a
conversion factor of $N$(H$_2$)/$T{_{\rm A}}$(CO)$\Delta v = 2 \times
10^{20} \mbox{ cm}^{-2} (\mbox{K} \mbox{ km s}^{-1})^{-1}$.  For the
\ion{H}{1}, we adopted a single value that characterizes the column
density across the galactic disk.  These values were drawn from maps
of galaxies reported in the literature.  Adopted values and the
corresponding references appear in Table \ref{h1src} along with the
orientation parameters and distances used in \citet{Helfer03}.  If no
\ion{H}{1} observations have been reported for the galaxies, we used a
value of 8 $M_{\odot}\mbox{ pc}^{-2}$, which corresponds to a surface
density of $1 \times 10^{21}$ cm$^{-2}$, typical for the stellar disks
of galaxies.


\begin{table}[h]
 \caption{\label{h1src}Adopted Galactic Parameters}
{\tiny
\begin{tabular}{ccccccl}
\hline\hline
Galaxy & Distance &Inc. & P.A. &
 $\Sigma_{\mathrm{HI}}$ & $\Sigma_{\mathrm{HI}}\Sigma_{*,t}^{1/2}$ & Reference \\
Name & (Mpc)& ($^{\circ}$)& ($^{\circ}$) &
 ($M_{\odot}/\mbox{pc}^{2}$) & ($M_{\odot}/\mbox{pc}^{2})^{3/2}$ & \\
\hline
 NGC 628  & 7.3 &  24 & 25 &  4.9  & 52  &   \citet{1992AnA...253..335K} \\
 NGC 1068 & 14.4 & 33 & 13 &  13.5 & 161 &\citet{1997ApnSS.248...23B} \\
 IC  342  & 3.9 &  31 & 37 &  4.0  & 59  &\citet{2001AJ....122..797C}\\
 NGC 2903 & 6.3 &  61 & 17 &  5.0  & 63  &\citet{1986AnAS...66..505W}\\
 NGC 3184 & 8.7 &  21 & 135 & 8.0  & 81  &---\\
 NGC 3351 & 10.1 & 40 & 13 &  5.0  & 56  &\citet{1989ApJ...343...94S}\\
 NGC 3368 & 11.2 & 46 & 5 &   4.0  & 46  &\citet{1988AnAS...73..453W}\\
 NGC 3521 & 7.2 &  58 & 164 & 8.0  & 75  &---\\
 NGC 3627 & 11.1 & 63 & 176 & 2.2  & 23  &\citet{1993ApJ...418..100Z}\\
 NGC 3726 & 17.0 & 46 & 10 &  7.0  & 62  &\citet{1986AnAS...66..505W}\\
 NGC 3938 & 17.0 & 24 & 0 &   7.5  & 67  &\citet{1982AnA...105..351V}\\
 NGC 4051 & 17.0 & 41 & 135 & 4.0  & 36  &\citet{2002ApJ...569..157W}\\
 NGC 4258 & 8.1 &  65 & 176 & 4.5  & 53  &\citet{1986AnAS...66..505W}\\
 NGC 4303 & 15.2 & 27 & 0 &   8.6  & 96  &\citet{1994AJ....107.1003C}\\
 NGC 4321 & 16.1 & 32 & 154 & 5.0  & 49  &\citet{2002ApJ...569..157W}\\
 NGC 4535 & 16.0 & 45 & 28 &  4.3  & 38  &\citet{1994AJ....107.1003C}\\
 NGC 4569 & 16.8 & 62 & 23 &  5.4  & 51  &\citet{1994AJ....107.1003C}\\
 NGC 4579 & 16.8 & 37 & 95 &  3.2  & 27  &\citet{1994AJ....107.1003C}\\
 NGC 4736 & 4.3 &  35 & 100 & 8.0  & 123 &\citet{2002ApJ...569..157W}\\
 NGC 4826 & 4.1 &  54 & 111 & 10.0 & 118 &\citet{1994ApJ...420..558B}\\
 NGC 5005 & 21.3 & 61 & 65 &  8.0  & 116 &---\\
 NGC 5033 & 18.6 & 62 & 170 & 8.0  & 62  &\citet{2002ApJ...569..157W}\\
 NGC 5055 & 7.2 &  55 & 105 & 6.3  & 72  &\citet{2002ApJ...569..157W}\\
 NGC 5248 & 22.7 & 43 & 110 & 8.0  & 93  &---\\
 NGC 5247 & 22.2 & 29 & 20 &  8.0  & 73  &---\\
 NGC 5457 & 7.4 &  27 & 40 &  6.3  & 81  &\citet{2002ApJ...569..157W}\\
 NGC 6946 & 5.5 &  54 & 65 &  7.8  & 95  &\citet{1986ApJ...308..600T}\\
 NGC 7331 & 15.1 & 62 & 172 & 8.0  & 87  &---\\
\hline
\end{tabular}
}
\end{table}

The stellar surface densities are determined using reduced $K$-band
images from the 2MASS Large Galaxy Atlas \citep{2mass-lga}.  We
adopted a constant $K$-band mass-to-light ratio of $M_L/L_K = 0.5
M_{\odot}/L_{\odot}$ \citep{m2l-bell}.  Our final sample consists of
only those galaxies with both 2MASS data and CO detections.  Of these,
22 galaxies have \ion{H}{1} surface densities in the literature and 6
galaxies do not.  The stellar and gas surface densities are corrected
to face-on values using the orientation parameters listed in
Table \ref{h1src}.

\begin{figure}[h!]
\begin{center}
\plotone{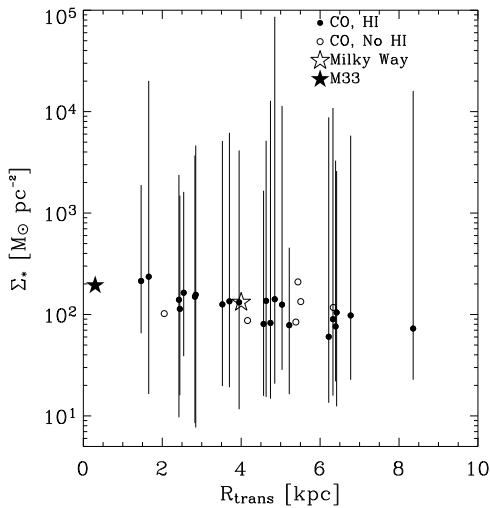}
\caption{\label{pressplot} Plot of the median stellar surface mass
density where $N$(H$_2)\approx N$(\ion{H}{1}) as a function of where
this surface density occurs in the galaxy.  For galaxies with measured
\ion{H}{1} densities (filled circles, 22 galaxies), the {\it range} of
stellar surface densities is plotted as an error bar, running between
the values of the 1st percentile and the 99th percentile of surface
density in the galaxy.  Galaxies without \ion{H}{1} measurements are
plotted as open circles (6 galaxies).  The transitions stellar surface
density is remarkably constant and has a mean value of $120\pm 10~
M_{\odot}\mbox{ pc}^{-2}$ for the 22 galaxies with both CO and
\ion{H}{1} data.  Points for the Milky Way (MW) and M33 are also plotted.}
\end{center}
\epsscale{1.0}
\end{figure}

We define the transition stellar surface density ($\Sigma_{*,t}$)
where $N$(H$_2$)=$N$(\ion{H}{1}), as the median value of the stellar
surface density at all positions for which $0.9
N(\mbox{\ion{H}{1}})\leq N(\mbox{H}_2)\leq 1.1 N(\mbox{\ion{H}{1}})$.
We associate this surface mass density with the transition radius
$R_{t}$ in the galaxy by finding the radius where the azimuthally
averaged stellar surface mass density $\overline{\Sigma_*}(R_{gal})$
equals the transition surface density, $\Sigma_{*,t}$.  In Figure
\ref{pressplot}, we plot the transition radius against the transition
stellar surface density, using the adopted distances in Table
\ref{h1src}.  We also list the values of $\Sigma_{\mathrm{HI}}\sqrt{\Sigma_*,t}$ are listed in
Table \ref{h1src} as a check since $P_{ext}$ depends directly on this quantity.
 
Figure \ref{pressplot} shows a remarkable constancy of $\Sigma_{*,t}$
expected if $N$(H$_2$)/$N$(\HI) is determined by pressure only; the
mean value is $120\pm 10~ M_\sun$ pc$^{-2}$.  Also plotted in the
figure is the range of $\Sigma_*$ measured for each individual galaxy.
Formally, the dispersion in the mean value of $\Sigma_{*,t}$, is only
40\%, yet the the range of $\Sigma_*$ to which the 2MASS survey is
sensitive in these galaxies varies by almost 3 orders of magnitude.
The scatter in $\Sigma_{\mathrm{HI}}\sqrt{\Sigma_{*,t}}$ given in
Table \ref{h1src} is
only 60\%.  The range of galactocentric distance where the transition
radius occurs varies by more than an order of magnitude.  Apparently,
the constancy of $\Sigma_{*,t}$ is not due to a small range in the
observed properties of the galaxies. The small scatter in
$\Sigma_{*,t}$ also suggests that various assumptions such as that of
a constant value of $h_*$ both within and among galaxies are
justified.

As a check, one can calculate the value of $\Sigma_{*,t}$ 
for the Milky Way, scaling the measured $\Sigma_*$ at
the distance of the Sun (35 M$_\sun$ pc$^{-2}$; Binney \& Merrifield
1998), and a radial scale length for the stars of 3 kpc (Spergel,
Malhotra \& Blitz 1996; Dehnen \& Binney 1998).  The transition radius
for the Milky Way occurs at a galactocentric distance of about 4 kpc
(Dame et al. 1993), or about 1.3 scale lengths inward of the Sun.
This converts to a stellar surface density of 132 M$_\sun$ pc$^{-2}$,
in good agreement with the determinations in other galaxies.  The
trend can also be checked in M33 using the data for $f_{mol}$ presented
in \citet{heyer-schmidt}, which gives $\Sigma_{*,t}=190 M_{\odot}\mbox{
pc}^{-2}$ where $\Sigma_{\rm{HI}}=\Sigma_{\mathrm{H2}}$ ($R_{gal}=300
\mbox{ pc}$).

There appears to be a small but significant decrease in the transition surface
density with radius, which may be due to a breakdown of the
assumptions of a constant mass-to-light ratio, $v_g$ and $h_*$ with
radius, or that pressure alone determines $N$(H$_2$)/$N$(\HI). 
A linear fit to the data in Figure 1 gives 
$\log \Sigma_* = (2.36 \pm 0.05) - (0.06 \pm 0.01)(R/\mbox{1 kpc})
M_{\odot}~\mbox{pc}^{-2}$ with errors given by the scatter in the data
around the trend.

\section{Discussion}
\label{discussion}

Figure 1 is consistent with the hypothesis that the mean hydrostatic
pressure determines the ratio of atomic to molecular gas at a given
radius in a disk galaxy.  The small scatter in the mean value of
$\Sigma_*$ suggests that globally, the pressure may be the {\it only}
important factor in determining the ratio of atomic to molecular gas.
But the variation in the hydrostatic equilibrium of a disk is expected to
be rather smooth.  Why then do galaxies show so much azimuthal variation
in the molecular gas, and by implication in the atomic-to-molecular gas
ratio (Helfer et al. 2003)?  Significant variations in the interstellar
pressure can result from a variety of causes such as spiral shocks and 
explosive events (e.g. supernovae).  Thus large pressure variations
can occur on all scales locally even if the mean hydrostatic pressure
varies only slowly.  Furthermore, even if the hydrostatic pressure
drives the ratio globally, locally the radiation field can be important
in determining how much of a molecular cloud can remain neutral
\citep{1999RvMP...71..173H}.
Thus, significant deviations from the mean molecular abundance can
be produced by both pressure and radiation variations.

Pressure or density?  The rate of formation of molecular gas is
thought to be dependent on the local gas density in the chemical reactions that
produce H$_2$ and CO.  Are we then using pressure as a
surrogate for the mean gas density
in determining $N$(H$_2$)/$N$(\HI) as a function of galactocentric
distance?  Pressure, as defined in Equations 1 and 2, is taken to be
$\rho_g \sigma_g^2$ where $\sigma_g$ includes both thermal and
turbulent
contributions.  Because $\sigma_g$ is measured to be typically
7 -- 8 \kms and the gas temperature of the cold gas layer where most
of the gas mass resides is typically $\lesssim$ 100 K, the thermal
pressure is only a small fraction of the turbulent pressure.
Observationally, $\sigma_g$ is constant
with radius (see \S 2), thus $P_{ext} \propto \rho_g$; variations in
pressure are effectively the same as variations in density.  We choose
to describe the functional dependence in terms of pressure rather than
density because it is directly measurable through Equation 2 (assuming
$h_*$ is known), whereas the density is inferred and not directly
measurable on galactic scales.  But it should be kept in mind that
given the measured constancy of $\sigma_g$, we cannot distinguish the
effects of pressure from those of density.

What pressure is implied by $\Sigma_{*,t}$?  In the Milky Way,
the value of $h_*$ is about 300 pc (Binney \& Merrifield 1998),
$v_g$ is about 7 \kmpers~(Dickey \& Lockman 1990), $\Sigma_g$ is
8.6 M$_{\sun}$~pc$^{-2}$, and $\Sigma_*$ is 132 M$_{\sun}$ pc$^{-2}$.
Using Equation \ref{approxpressure}, $P_{ext}/k$ = 1.5 $\times 10^4$
cm$^{-3}$ K after correcting for helium.  This value is still an order
of magnitude below the mean internal $P_{int}/k$ $\sim3$ $\times 10^5$
cm$^{-3}$ K for GMCs that have typical surface densities of $\sim$
100 M$_{\sun}$~pc$^{-2}$ \citep{1993prpl.conf..125B}.  Because
$P_{int}\gg P_{ext}$, GMCs that survive for more than a crossing time, $\sim
10^7$ y, must be self-gravitating.  Using Equation 2 to scale $P_{ext}$
to the inner regions of disks suggests that GMCs are self-gravitating
over nearly the entire disk.

What are the implications for understanding star formation on galactic
scales?  If the global atomic--molecular transition is governed by
pressure across a wide range of galaxies, it will be possible to
develop a perscription for 
star formation on global scales that is physically well motivated.  
Non-thermal radio-continuum is
tightly correlated with the far IR emission in galaxies
\citep[e.g.][]{1992ARA&A..30..575C}, implying that the radio-continuum
is a good extinction-free indicator of the star formation rate in
spiral galaxies.  \citet{2002A&A...385..412M} have shown that for a
sample of 180 spiral galaxies, the ratio of radio-continuum to CO
emission is constant to within a factor of 3, suggesting that the star
formation efficiency of molecular clouds averaged over galactic scales
is constant at about 3.5\%.  Therefore, if the relationship between
pressure and $N(\mbox{H}_2)/N(\mbox{\ion{H}{1}})$ (i.e the function
$f$ in Eqs. 1 and 2) can be determined for all galaxies, or if the
variation in $f$ can be found for different galaxy types, then it will
be possible to determine the star formation rate in galaxies by
measuring the stellar and gas surface densities only.  Furthermore, 
it will be possible to obtain reliably the star
formation rate from simulations, since the turbulent gas pressure can
be directly calculated. In addition, if
the variation in $f$ can be measured for galaxies of low metallicity,
then determining the star formation rate can be extended to high $z$.
The measurement of $f$ will be the subject of
a future paper.

We suggest, then, that GMCs can form from either preexisting
atomic or molecular gas depending on the dominant state of the
diffuse interstellar medium at a particular radius in a galactic disk.
That dominant state is determined by the hydrostatic pressure,
modified by local perturbations such as density waves, supernova
remanants, etc.

\acknowledgements 
We thank an anonymous referee and Tony Wong for comments which
improved the clarity of the paper.  ER's work was supported, in part,
by a NASA Graduate Student Research Program Fellowship. This work is
partially supported by NSF grant 0228963 to the Radio Astronomy
Laboratory at UC Berkeley.

\end{document}